\documentclass[
 aip,
 amsmath,amssymb,
preprint,
]{revtex4-1}

\usepackage{graphicx}
\usepackage{dcolumn}
\usepackage{bm}

\usepackage[utf8]{inputenc}
\usepackage[T1]{fontenc}
\usepackage{mathptmx}
\usepackage{etoolbox}
\usepackage[dvipsnames]{xcolor}

\makeatletter
\def\@email#1#2{%
 \endgroup
 \patchcmd{\titleblock@produce}
  {\frontmatter@RRAPformat}
  {\frontmatter@RRAPformat{\produce@RRAP{*#1\href{mailto:#2}{#2}}}\frontmatter@RRAPformat}
  {}{}
}%
\makeatother
\begin{document}

\preprint{}

\title[]{Direct frequency comb cavity ring-down spectroscopy:\\Enhancing Sensitivity and Precision}
\author{R. Dubroeucq}
\affiliation{ 
Université Rennes, CNRS, IPR (Institut de Physique de Rennes)-UMR 6251, F-35000 Rennes, France.
}

\author{D. Charczun}%
\affiliation{ 
Institute of Physics, Faculty of Physics, Astronomy and Informatics, Nicolaus Copernicus University in Toru\'{n}, ul. Grudziadzka 5, 87-100 Toru\'{n}, Poland 
}

\author{P. Mas\l owski}
\affiliation{ 
Institute of Physics, Faculty of Physics, Astronomy and Informatics, Nicolaus Copernicus University in Toru\'{n}, ul. Grudziadzka 5, 87-100 Toru\'{n}, Poland 
}
\email{pima@fizyka.umk.pl}
\author{L. Rutkowski}
\affiliation{ 
Université Rennes, CNRS, IPR (Institut de Physique de Rennes)-UMR 6251, F-35000 Rennes, France.
}
 \email{lucile.rutkowski@univ-rennes.fr}
\date{\today}

\begin{abstract}
We present a novel approach to cavity ring-down spectroscopy utilizing an optical frequency comb as the direct probe of the Fabry-Perot cavity, coupled with a time-resolved Fourier transform spectrometer for parallel retrieval of ring-down events. Our method achieves high spectral resolution over a broad range, enabling precision measurements of cavity losses and absorption lineshapes with enhanced sensitivity. A critical advancement involves a stabilization technique ensuring complete extinction of comb light without compromising cavity stabilization. We demonstrate the capabilities of our system through precision spectroscopy of carbon monoxide rovibrational transitions perturbed by argon.
\end{abstract}

\maketitle

\section{\label{sec:level1}Introduction}

Cavity ring-down spectroscopy (CRDS) stands as a cornerstone in modern spectroscopic techniques, thank to its unparalleled sensitivity and unique attributes such as calibration-free operation and immunity to light intensity fluctuations. Traditionally reliant on continuous wave (cw) lasers for operation, CRDS has seen recent advancements aimed at boosting acquisition rates and spectral density, achieving scanning speeds up to 1 THz/s through adaptive spectral sampling based on innovative methodologies like electro-optic modulation \cite{long2014,long2015}, Mach-Zehnder modulators\cite{burkart2015,votava2022} and frequency-swept synthesizers\cite{gotti2020}.
However, existing cw-laser methods entail sequential acquisition of spectral elements, making experimental data more sensitive to time variation of important thermodynamic parameters and generally longer to acquire. Therefore there is a  demand for broadband instruments capable of providing multiplexed spectra with similar spectral resolution and absorption sensitivity. 

Historically, efforts to achieve broadband CRDS date back to pioneering works such as those by Engeln \& Meijer\cite{engeln1996} utilizing pulsed dye lasers and step-scan time-resolved Fourier transform spectrometers. However, limitations in resolution and acquisition time persisted, hindering broader applications.. Subsequent techniques, including ring-down spectral photography by Scherer\cite{scherer1998} and time-resolved spectra acquisition by Czyżewski et al.\cite{czyzewski2001}, demonstrated various approaches to achieve time-resolved measurements but with limited performances in terms of resolution and sensitivity. Notably, Thorpe et al.\cite{thorpe2006} proposed for the first time employing an optical frequency comb as the light source for broadband CRDS, to revisit the spectral photography approach with enhanced sensitivity.

In 2022, two new developments in the near-infrared  laid the groundwork for enhanced acquisition speeds and resolution by combining CRDS with dual-comb interferometric detection\cite{lisak2022} and time-resolved Fourier transform spectroscopy\cite{dubroeucq2022}. Lisak \textit{et al.} \cite{lisak2022} relied on  dual-comb CRDS with electro-optics combs pumped from the same seed cw laser. The high acquisition speed of a single time-resolved interferogram (5 $\mu$s) allowed several interferograms to be recorded during a single cavity decay to yield the decay time-dependence. The setup demonstrated a sensitivity of 3×10$^{-8}$ cm$^{-1}$, with a spectral point spacing of 1 GHz, a simultaneous spectral coverage of 22 GHz, for an acquisition time of 1 s and a cavity finesse around 20,000. Dubroeucq \& Rutkowski\cite{dubroeucq2022} recently demonstrated multiplexed CRDS using an optical frequency comb as the direct probe and a fast-scanning time-resolved Fourier transform spectrometer, facilitating quantitative measurements of CO$_2$ and H$_2$O concentrations in ambient air. The optical frequency comb modes were directly locked to the cavity resonances using the Pound-Drever-Hall (PDH) technique\cite{black2001}. The spectrometer allowed measuring spectra covering 1.2 THz with a resolution of 600 MHz, achieving a noise equivalent absorption of 1.5×10$^{-8}$ cm$^{-1}$. The main challenge was to decouple the ring-down signals from the stabilization error signals. Indeed, in this first proof-of-concept, the comb light reflected from the cavity was retrieved to generate the PDH error signal. An acousto-optic modulator (AOM) would then shut off the comb light to trigger decay events, while simultaneously shutting off the error signal. The latter meant that a relative comb-cavity passive stability at the time scale of the shut-off duration was required to perform the measurement. This limited the AOM shut-off duration, which in turn prevented using high-finesse cavities characterized by long ring-down times. Very recently, a similar approach was proposed by Liang \textit{et al.}\cite{liang2024}, where no active synchronization of the FTS movement and ring-down events was performed. The experiment rather relied on the performance of the FTS translation stage, whose speed was constant enough to allow demodulation of the ring-down events during post-processing. The demonstration was performed directly in the mid-infrared range and yielded the broadest spectral range measured simultaneously using cavity-enhanced direct frequency comb spectroscopy. 

The present work builds upon the two approaches published in 2022\cite{dubroeucq2022,lisak2022} to achieve high-sensitivity and high-resolution cavity ring-down spectroscopy utilizing an optical frequency comb and time-resolved Fourier transform spectrometry. Here, we address key limitations identified in prior proof-of-concept experiments, enhancing sensitivity and precision by orders of magnitudes while facilitating quantitative measurements of gas concentrations. The sensitivity limitations are exceeded by a fundamental redesign of the frequency  stabilization method enabling the extinction of the comb light over a long duration without loosing the comb-cavity lock. A cw-laser is introduced to act as an intermediary between the comb and the cavity, effectively making the comb-cavity stabilization scheme insensitive to shut-off of the probe beam. We achieve ring-down spectroscopy using a 20,000 finesse cavity and demonstrate an order of magnitude improvement of the absorption sensitivity compared to the state-of-the-art systems\cite{lisak2022,dubroeucq2022}. These performances allowed retrieving high signal-to-noise ratio absorption spectra of CO mixed with Ar over a broad coverage and confirming the influence of speed-dependent effects on the absorption line profiles. Our findings offer promising prospects for applications spanning from environmental monitoring to fundamental spectroscopic research.

 \section{Fourier-transform cavity ring-down spectroscopy}
\begin{figure*}
\centering
\includegraphics[width=1\linewidth]{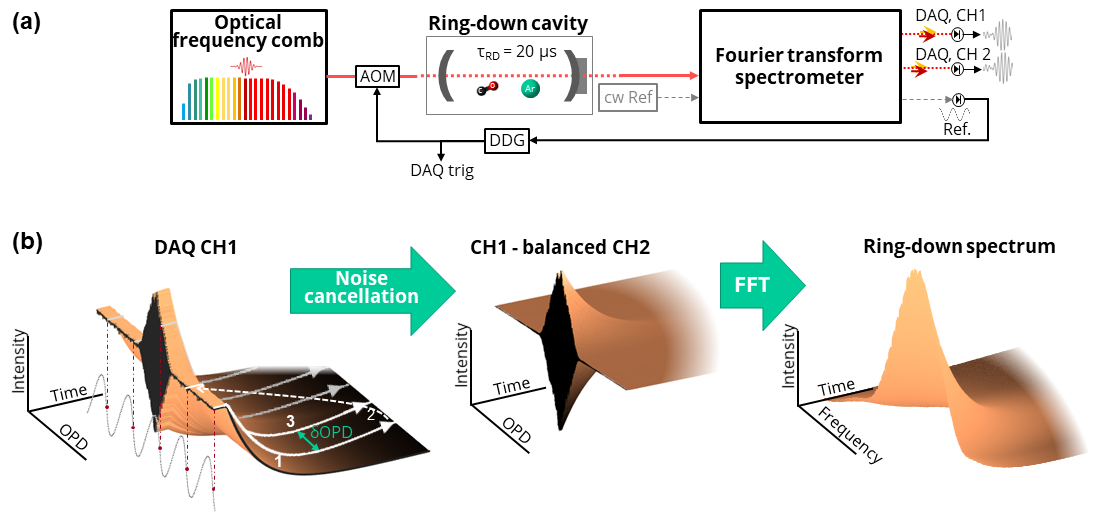}
\caption{(a) Simplified schematic of the Fourier transform CRDS setup. The light emitted by the comb source propagates through an acousto-optic modulator (AOM) before reaching the optical cavity, enclosed in a tight cell filled with a gas mixture, and the transmitted cavity light is analyzed using a fast-scanning Fourier transform spectrometer (FTS). A frequency stabilized cw reference laser propagates in the FTS on a path parallel to the comb beam, yielding a sinewave interferogram used to trigger the AOM shut off and triggers the data acquisition (DAQ) system via a digital delay generator (DDG). The two optical outputs of the comb beam are recorded simultaneously using two DC-coupled detectors connected to the two channels of the DAQ (CH1,2). (b) Principle of the acquisition and data processing. The signal measured at one of the comb optical outputs contains a series of cavity decays separated in optical path difference (OPD) by a constant step given by the reference laser wavelength. The subtraction of the two FTS outputs allows suppressing the intensity noise affecting the decay amplitude. Fourier transformation of the interferogram along the OPD axis yields frequency-sorted cavity decays.}
\label{fig:principle}
\end{figure*}
\subsection{Mulitplexed CRDS}
Cavity ring-down spectroscopy operates in the time domain, requiring a methodical approach to measure light-intensity changes over time. In its cw-laser implementation, CRDS relies on a fast shut-off of the laser light when its frequency matches one of the cavity resonance modes. The ring-down measurement is performed after the shut-off of the injected light on the cavity transmitted intensity: the intra-cavity power decreases exponentially with time, with a characteristic ring-down time $\tau_{RD}$ set by the total loss rate of the cavity which in realistic scenarios of absorption measurements is frequency-dependent. When the optical cavity is injected with an optical frequency comb with repetition rate $f_{rep}$ and offset frequency $f_0$ matched to the cavity modes, the intra-cavity field contains all the comb frequency components at once, all subject to different cavity-loss rates. Untangling the spectral variation of the cavity ring down-time therefore requires a detection able to sort the spectrum while measuring the time-variation of the cavity transmission. 

Time-resolved Fourier transform spectrometry (FTS) based on a Michelson interferometer\cite{gerwert1999,picque2000} is an effective solution\cite{dubroeucq2022}, at the necessary and sufficient condition that the cavity decays are synchronized with the interferometer movement for accurate analysis\cite{griffiths2007}. A simplified schematic of the experimental approach is given in Fig.\ref{fig:principle}(a). The optical frequency comb is frequency matched to the cavity modes and an acousto-optic modulator is inserted in the optical path between the two to act as an instantaneous light switch. The comb light transmitted through the cavity is directed to a fast-scanning FTS, which is similar to previous designs \cite{foltynowicz2011, nishiyama2020, dubroeucq2022}. A frequency-stable HeNe laser beam propagates in the interferometer on a path parallel to the comb beam direction. The reference sinewave interferogram it provides is later used to calibrate the variation of the optical path difference (OPD) of the interferometer. In case of a time-resolved measurement, this sinewave is used to trigger the time-dependent event with a constant OPD step and the interferogram acquisition. Here, the time-dependent event is the cavity multi-exponential decay and it is measured at the two optical outputs of the interferometer (CH1 and CH2, in Fig.\ref{fig:principle}(a)). As it is already commonly known in FTS\cite{foltynowicz2011}, these two signals share the same intensity noise but the interferograms are out-of-phase due to an odd or even, respectively, number of reflections on the interferometer path. It is worth mentioning that in the case of a time-dependent measurement, analog auto-balanced detection approaches are ineffective: the interferograms contain very low frequencies and the electronic feedback would bring the difference to zero. Instead, the two signals are measured independently using two distinct low-noise photodiodes and two channels of the acquisition card.

\subsection{Acquistion process}
The data acquisition process and processing are schematically depicted in Fig. \ref{fig:principle}(b). The objective is the acquisition of a three-dimensional interferogram, where the third dimension accounts for the cavity decay time. The OPD scan is continuously scanned at a constant speed and every occurrence of the reference intensity (grey dotted line below the 3D interferogram) crossing a predefined threshold (red circular markers) triggers the AOM shut-off. The first channel of the data acquisition system (DAQ, CH1) simultaneously records a cavity decay (white plain curve) with a time only set by the acquisition card and photodiode bandwidths. After acquiring a single event, the system waits for the next reference crossing to start a new decay at a different OPD, and repeats the same procedure to reach the total OPD which sets the spectral resolution of the final spectrum. The interferometer never stops scanning, therefore the OPD varies along a single cavity decay, but this does not impact the final analysis as long as the OPD step is constant at any time after the AOM trigger (i.e. as long as the OPD variation speed is constant). As in cw laser-based CRDS, single decays are inherently immune to intensity noise as the measurement is performed when the laser is off. However, many decays are necessary to record the full interferogram, and the initial laser intensity varies shot-to-shot. This intensity noise is generally observable on the baseline of the interferograms recorded at the FTS output and would be transferred to the spectrum if not cancelled in post-processing. To overcome this limitation, we make use of the two acquired interferograms and subtract one from the other after a light correction of the alignment variation and photodiode gain difference, similarly to the procedure described in our previous work\cite{dubroeucq2022}. The decay spectrum is obtained after calculating the magnitude of the fast Fourier transformation (FFT) of the decay interferogram along the OPD axis, yielding the spectral distribution of the cavity response.

\section{Experimental implementation}
\subsection{Cavity ring-down spectrometer}
The CRDS setup follows the description given in Fig.\ref{fig:principle}(a). The optical frequency comb was emitted by a Erbium:fiber mode-locked laser (MenloSystems) covering the 1.5-1.6 $\mu$m spectral range with a 250 MHz repetition rate ($f_{rep}$) and an average power of 250 mW. The laser output was collimated in fiber to an AOM operating at $f_{AOM}$ = 200 MHz, enabling a light shut-off with a response time faster than 10 ns. The comb light transmitted by the AOM was collimated into free-space and subsequently matched to inject the Fabry-Perot enhancement cavity efficiently. The ring-down cavity was composed of two high reflectivity mirrors (R>99.985\%) spatially separated by a distance of 60 cm to yield a cavity free spectral range (FSR) matching $f_{rep}$. The empty cavity finesse was 21,400(60) at 1545 nm, which translates to an empty cavity ring-down time of 12.7 $\mu s$ when considering the cavity length. One of the mirrors was mounted on a piezo transducer (PZT) to allow active control of the cavity length. The entire cavity was enclosed in a tight cell where the gas mixture and pressure could be controlled. The transmission was collimated again in fiber to reach the fast-scanning FTS, mounted in a twice-folded configuration\cite{nishiyama2020}. The translation stage was a 120 cm long Aerotech ACT165DL-1200, which was continuously operated at a displacement speed of 1 mm/s,  travelling over a total distance of 31.6 cm in 5 min 16 s (yielding a total OPD of 120 cm and a nominal resolution of 250 MHz). The reference cw laser used to calibrate the OPD and trigger the acquisition was a frequency stabilized He-Ne laser emitting at 633 nm. The two optical outputs of the comb beam were monitored using two identical Newport photodetectors (2053-FS, set at 5 MHz bandwidth) and acquired using a National Instrument PCI5922 acquisition card in Multiple-Record acquisition mode (sampling rate of 5 MS/s, alias-free bandwidth of 2 MHz, 20 bit resolution). 350 points were measured for each decay and the acquisition OPD step was equal to the He-Ne laser wavelength. 

 \begin{figure*}
     \centering
     \includegraphics[width=1\linewidth]{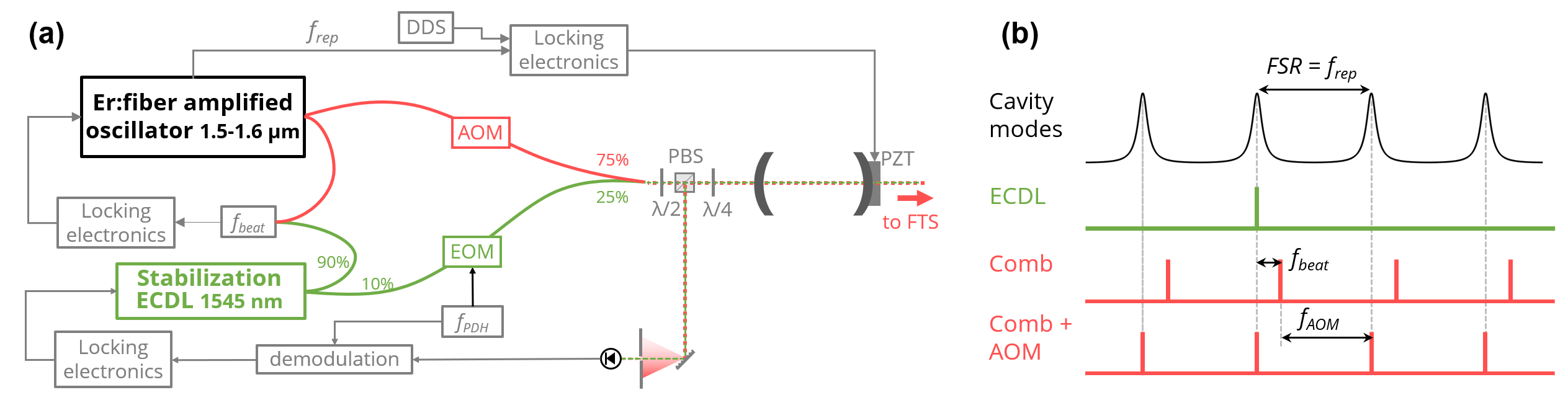}
     \caption{Frequency stabilization of the spectrometer: (a) schematic setup and (b) frequency domain matching. The ECDL is phased modulated at $f_{PDH}$ in an electro-optic modulator (EOM) to be locked to the ring-down cavity using the Pound-Drever-Hall (PDH) locking technique. The cavity reflected beam is picked up using a set of wave plates ($\lambda$/2,4) and a polarizing beam splitter (PBS). The optical beatnote $f_{beat}$ between the ECDL frequency and the closest comb mode is stabilized to a radiofrequency reference so that when summed with $f_{AOM}$, the frequency shift experienced by the comb in the AOM, it matches the cavity free spectral range $FSR$ and the laser repetition rate $f_{rep}$.  The absolute frequency stability is obtained by locking $f_{rep}$ to a tunable radiofrequency generated by a direct digital synthesizer (DDS) via a feedback to a piezo actuator (PZT) controlling the cavity length. \textit{Not shown:} the comb offset frequency is stabilized to a constant value using commercial electronics.}
     \label{fig:setup}
 \end{figure*}
 
\subsection{Stabilization scheme}

To accommodate a light shut-off lasting 70 $\mu s$ while actively maintaining the comb-cavity frequency matching, an intermediary cw laser was used to stabilize the spectrometer, as is depicted in Fig.\ref{fig:setup}(a). This intermediary was an external cavity diode laser (ECDL, RIO) emitting at 1545 nm. A small fraction of its total power was combined with the infrared comb to propagate to the cavity. The ECDL frequency was locked to one of the cavity modes using the Pound-Drever-Hall locking technique\cite{black2001}. As shown in Fig.\ref{fig:setup}(b) which depicts the frequency matching, the ECDL mode followed any frequency fluctuation of the cavity resonant mode. The remaining power of the ECDL was overlapped with a small part of the initial comb. The beatnote between the ECDL and the closest comb mode was stabilized with a high bandwidth at $f_{beat}$ = 50 MHz by acting on the comb intra-cavity actuators (namely an EOM and a PZT). The value of $f_{beat}$ was chosen so that when added to the frequency shift $f_{AOM}$ = 200 MHz induced by the AOM placed on the comb path, the comb modes would match the resonant modes of the ring-down cavity. This locking loop allowed the comb power to be entirely shut off without disrupting the stabilization of the system, as the AOM action impacted none of the error signals. It also enabled the comb modes to match the cavity at all times so that the comb light was directly resonant in the cavity when it was not shut off. This enabled the perfect synchronization of the decays with the FTS movement. In addition, the continuous comb-cavity matching was important in optimizing the overall measurement duty-cycle: 44\% of the measurement time was effectively dedicated to the measurement of the cavity decay while the remaining 56\% were used to reach a stationary intra-cavity build-up while waiting for the next reference trigger.

The absolute frequency stabilization was achieved by locking the offset frequency, $f_0$ - measured by the built-in \textit{f-2f} interferometer, to a constant value of 19.6 MHz and stabilizing the repetition rate to a tunable radiofrequency generated using a direct digital synthesizer (DDS). The latter stabilization loop was closed by feeding back the control signal to the cavity PZT to stabilize the cavity FSR and, in turn, $f_{rep}$. The comb parameters $f_{rep}$ and $f_0$, as well as $f_{beat}$ were continuously monitored during the acquisition using a frequency counter with a 1 Hz integration time. The monitoring results yield the following frequencies: $f_{rep} = 250,272,859.000(7)$ Hz, $f_{0} = 19,600,000(1.3)$ Hz and $f_{beat} = 50,366,100(7)$ Hz. All radiofrequency electronics were referenced to a hydrogen maser signal available in the laboratory via a fiber link to the Borowiec observatory\cite{sliwczynski2013,morzynski2015}.


\subsection{Ring-down spectrum processing}
\begin{figure}
    \centering
    \includegraphics[width=1\linewidth]{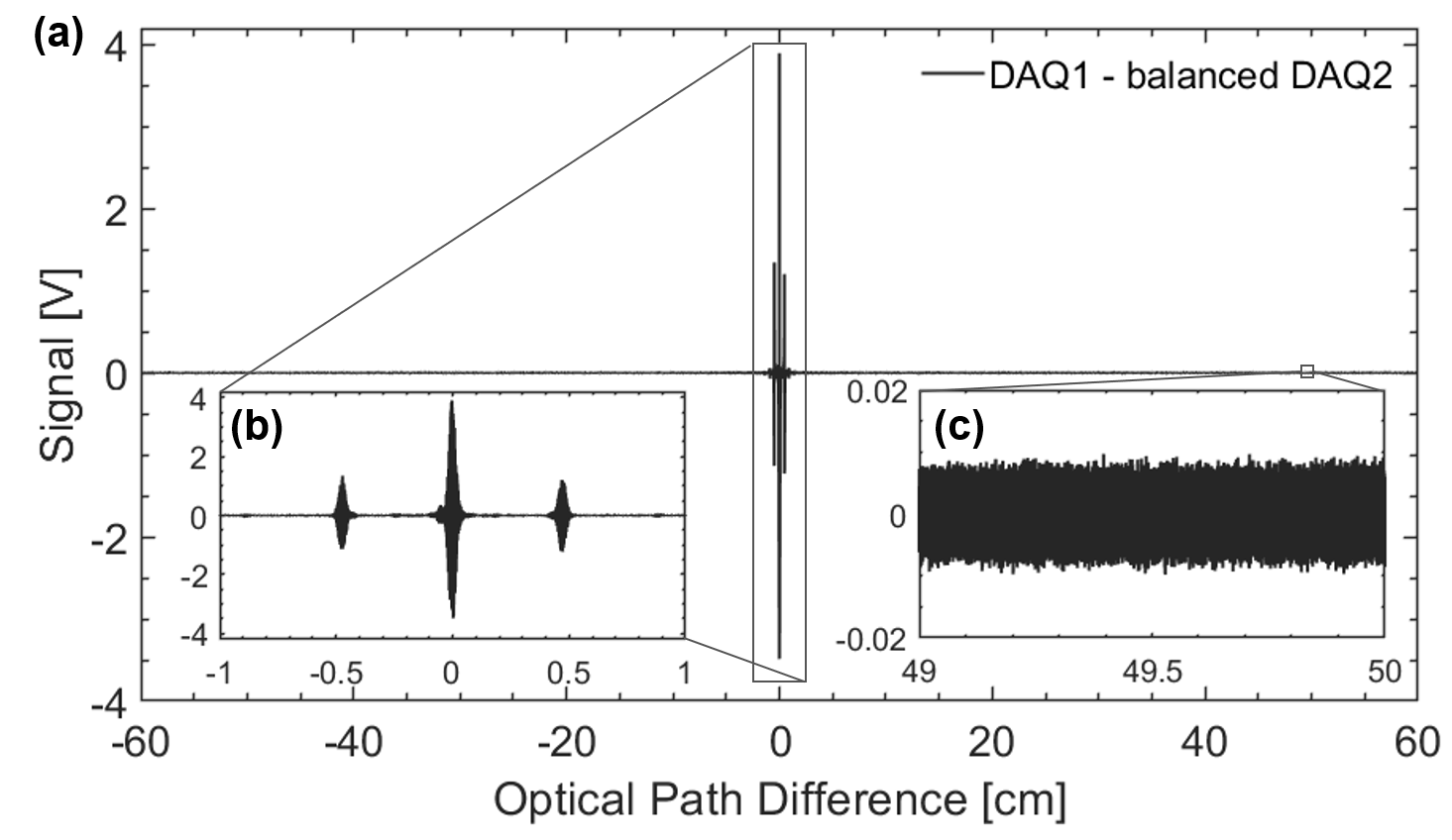}
    \caption{First interferogram obtained after the trigger signal, where the noise has been reduced by the balancing procedure: (a) full interferogram, (b) zoom around the centre burst, and (c) zoom at the end of the interferogram where most of the signal is noise. }
    \label{fig:interf}
\end{figure}

Figure \ref{fig:interf}(a) shows a typical interferogram obtained at the first time-domain point after the reference trigger (i.e. just at the beginning of the intensity decay at the cavity transmission). The subtraction between the two DAQ channels has been optimized so that the continuous component of the interferogram is zero. Panel (b) shows details of the center burst, peak-to-peak of which is equal to 7.4 V. The interferogram standard deviation reduces as the considered data points are further away from the center burst, to reach 5 mV at the edges of the interferogram, as shown in panel (c). This value is consistent with the specified noise performance, gain and bandwidth of the two Newport photodiodes used at the FTS outputs.

\begin{figure}
    \centering
    \includegraphics[width=1\linewidth]{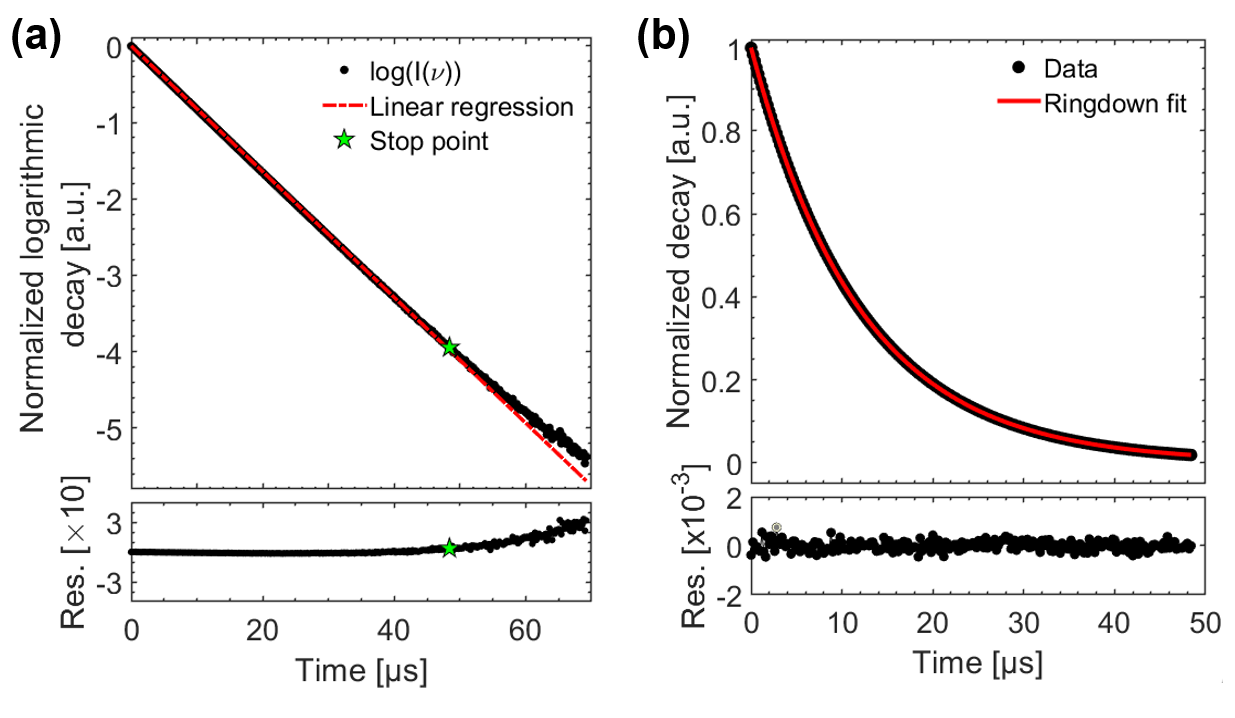}
    \caption{(a) - Logarithm of the ring-down at 1581 nm (black circles) plotted with best linear fit (red dashed line) in the top panel, and fit residuum in the lower panel. The green stars in both panels represent the point where the decay signal becomes equal to the noise. (b) - Linear decay data points cropped from (a) (black circles) and exponential decay model with fitted ring-down time and amplitude (red solid line). Lower panel shows the final fit residuum. }
    \label{fig:RD}
\end{figure}

The same processing was applied to all interferograms acquired at different delays after the reference triggers. Following the steps shown in Fig.\ref{fig:principle}(b), the cavity exponential decays are sorted in their respective spectral element. The time-resolved signals are then single exponential decays if the spectral resolution is high enough to assume a constant loss rate over a spectral element and if the acquisition is stopped before the signal reaches the noise level. The former condition can be fulfilled as one can tune the spectral resolution by changing the total OPD measured. The latter however is intrinsic to FTS measurement: all DC components are cancelled when the balancing procedure is performed, and therefore taking the magnitude FFT of the interferogram yields a distortion of the decay. This can easily be overcome by cutting the data set before the signal reaches the noise level.  One of these decays is shown in Fig.\ref{fig:RD}, selected at 1581 nm (6325 cm$^{-1}$). Panel (a) shows the natural logarithm of the signal retrieved at this spectral element, normalized to 1, together with a linear model with fitted characteristic time. The residuum of the fit is shown in the lower panel and clearly shows that the linear model only accounts for the first part of the data and not the last points when the residuum deviates from 0. To avoid bias due to this numeric artifact, the data is cropped so that only the points with signal higher than the noise value are kept and this limit is evaluated for each decay independently as it depends on the signal-to-noise-ratio, noise and characteristic time of the considered spectral element. In the case of the decay shown in Fig.\ref{fig:RD}, only the first 50 $\mu s$ of the signal were kept for the final ring-down fit. Panel (b) shows the linear data with an exponential decay fit where the ring-down time and the amplitude were adjusted, and the fit residuum is shown below. The residuum does not exhibit any particular structure and is characterized by a standard deviation of $1.9\times10^{-3}$, yielding a signal-to-noise ratio for this decay of 5278. The ring-down time $\tau_{RD}$ of this spectral element was 12.0 $\mu s$, with an error retrieved from the fit equal to 2 ns, far below the time uncertainty of the acquisition card (1 / 5 MHz = 250 ns). 

All spectral elements underwent the same procedure to obtain the spectral variation of the cavity ring-down time.

\color{black}

\section{High finesse cavity ring-down spectroscopy of CO in Ar }
\subsection{Sample conditions and retrieved spectrum}
\begin{figure}
    \centering
    \includegraphics[width=1\linewidth]{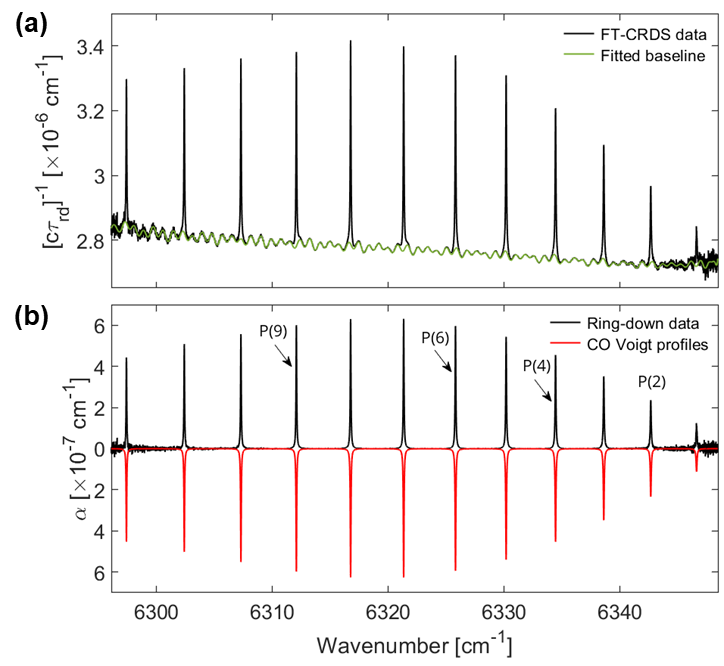}
    \caption{Ring-down spectrum retrieved from the multiplexed CRDS measurements, averaged 15 times. (a) Spectrum obtained after the first data processing (black curve), plotted together with the fitted baseline including a low polynomial functions and a set of sinewaves modelling optical fringes (green curve). (b) Absorption spectrum retrieved after baseline subtraction (black curve) compared with a synthetic CO absorption spectrum calculated using Voigt profiles and spectroscopic parameters from the HITRAN database. The lines pointed by arrows are the ones selected for further spectroscopic analysis.}
    \label{fig:full_spec}
\end{figure}

The spectrometer was applied to the retrieval of high signal-to-noise ratio of the absorption spectrum of CO diluted in a buffer of Ar. In addition to being a molecule highly relevant to atmospheric studies, CO is broadly considered to be a good prototype molecule to test quantum calculations of molecular lineshapes\cite{kowzan2017,kowzan2020jqsrt, kowzan2020pra} and the test bed for high order collisional physics\cite{reed2023}. Here, 250 ppm of CO was mixed with argon at a total pressure of 534(2) Torr, measured using a 1000 Torr MKS Baratron (resolution of 10 mTorr and 0.25\% uncertainty). The gas cell was tightly closed during the measurement, and no gas flow was used. The gas cell was not controlled in temperature, which was therefore varying a little during the experiment. The gas temperature considered in the analysis was 295(2) K. The intra-cavity pressure was measured before and after the measurement to ensure its stability. The ring-down acquisition was performed with a nominal spectral resolution of 250.27 MHz, just above the comb repetition rate so that the modes were not resolved by the FTS, and 15 successive datasets were recorded. Averaging of the datasets was performed in the frequency domain (hence after FFT) but before the ring-down fitting step. The spectrum of the inverse of the ring-down time multiplied by the speed of light $c$ is plotted in Fig.\ref{fig:full_spec}(a). As in traditional CRDS, this spectrum is the sum of the intra-cavity absorption - the CO lines are clearly visible - with a background set by the cavity mirror losses and optical fringes. The baseline is first fitted independently with a polynomial function and a set of sinewaves. To remove the effect of the absorption lines, a simple spectroscopic model relying on Voigt profiles, the spectroscopic parameters listed in the HITRAN database\cite{gordon2022} (with collisional broadening parameters corrected by a constant factor to account for the Ar buffer instead of $N_2$ as listed in HITRAN). The resulting baseline is subtracted from the data, which is shown in Fig.\ref{fig:full_spec}(b) in black together with the Voigt profile model (red curve, inverted for clarity). In the center part of the spectrum between $6315$ and $6335$ cm$^{-1}$, the noise equivalent absorption is equal to $1.1\times 10^{-9}$ cm$^{-1}$, which yields a figure of merit of $9.6\times 10^{-10}$ cm$^{-1}$Hz$^{-1/2}$ per spectral element when normalized by the total acquisition time.

\subsection{Lineshape analysis of the CO-Ar molecular system}
\begin{figure}
    \centering
    \includegraphics[width=1\linewidth]{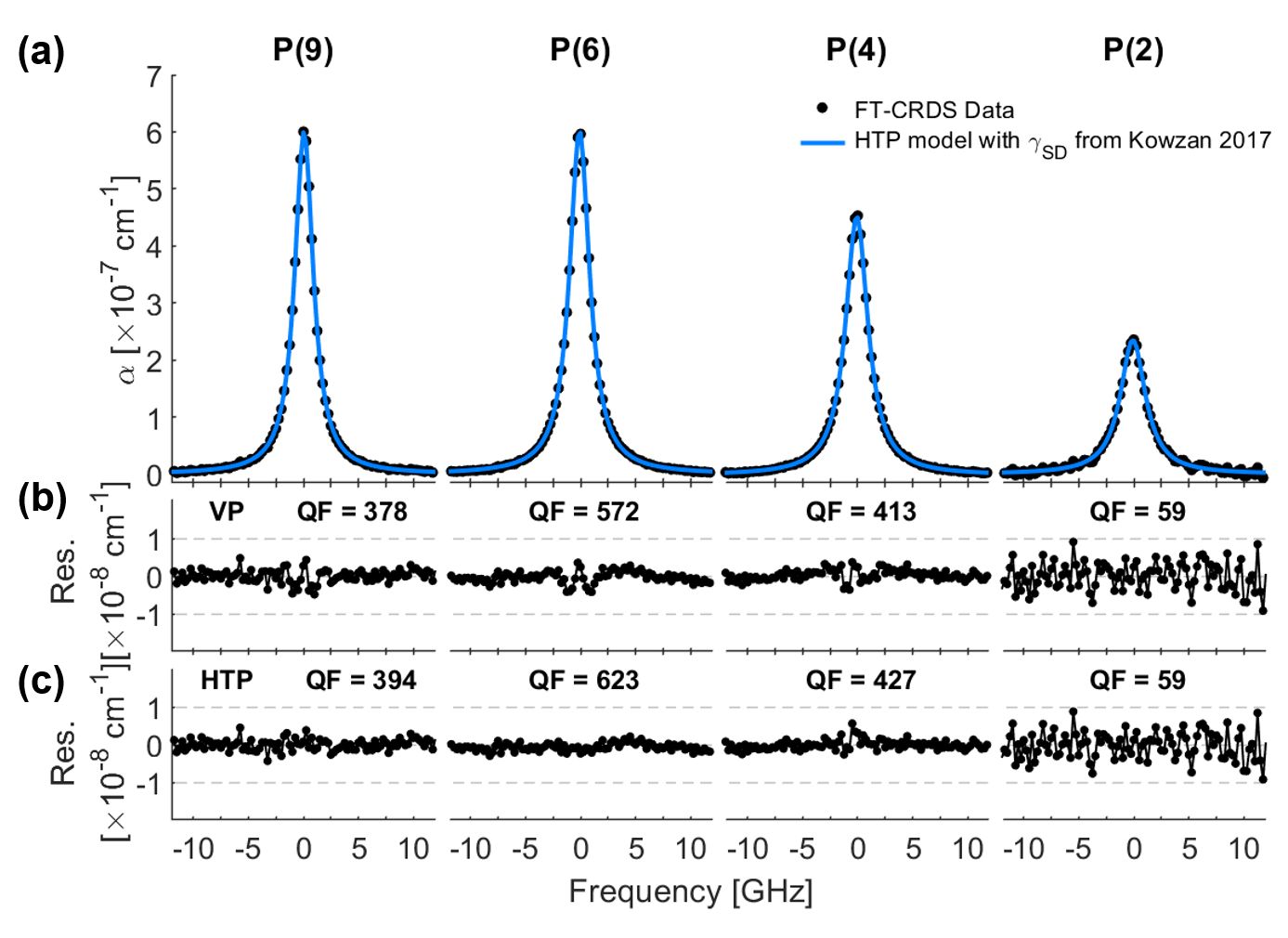}
    \caption{ (a) Measured absorption lines (back circular markers) plotted together with simulated lines calculated using the Hartmann-Tran profile (HTP, blue curve) with speed-dependent parameters from Kowzan \textit{et al.}\cite{kowzan2017}. (b) Residua of the data fitted using Voigt profile models, see text for discussion. (c) Residua of the data fitted using the reduced HTP profiles. The quality factors (QF) indicated in each residuum panel are calculated as the maximum absorption of the considered line divided by the root mean square error of the residuum\cite{cygan2012}.}
    \label{fig:HTP}
\end{figure}

\begin{figure}
    \centering
    \includegraphics[width=1\linewidth]{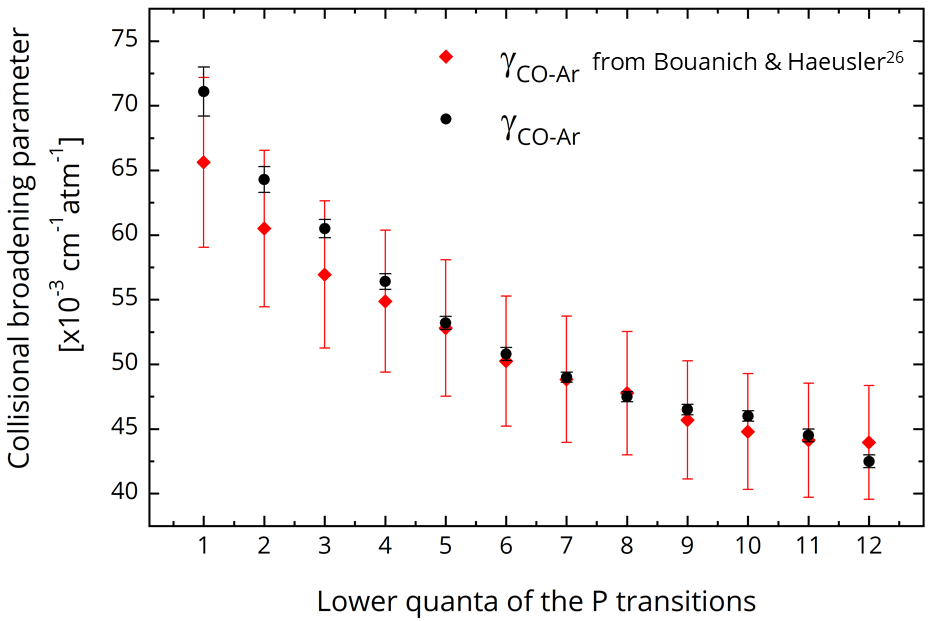}
    \caption{Pressure broadening parameters for CO in Ar retrieved from the multiplexed CRDS spectrum as a function of the lower quanta of the P transition: the present work is shown in black circular markers, and previously available data are plotted in red diamond markers\cite{bouanich1972}.}
    \label{fig:gamma}
\end{figure}

To perform the spectroscopic analysis of the spectrum, we focused on the lines that were already studied in previous works\cite{kowzan2020jqsrt,kowzan2020pra}: the P(9), P(6), P(4) and P(2) lines, plotted in black circular markers in \ref{fig:HTP}(a). As a first step, to compare the results with previous studies, these absorption lines were modelled using Voigt profiles with linestrengths fixed to their HITRAN values, and where the pressure broadening, the temperature and the pressure were fitted to the experimental data. The retrieved temperature was 295.2(1.5) K, and the CO concentration was found equal to 244.6(2.1) ppm. The retrieved broadening parameters (half-width at half maximum) for CO in Ar are shown in Fig.\ref{fig:gamma} (black circular markers). The error bars of our results were estimated based on the fit uncertainty and error propagation from the experimental parameters. The value plotted has also been corrected for the broadening contributions of the CO-CO collisions, for which the parameters listed in HITRAN were considered. The results of this work are compared with the parameters retrieved from a previous study\cite{bouanich1972} (red diamond markers): both datasets agree within error bars.

The residuals of the fits are shown in Fig.\ref{fig:HTP}, where a systematic deviation is visible in the line centers. Investigations showed that this remaining structure in the residuum could not be explained by an error in the retrieved temperature and concentration, nor by an incorrect baseline estimation. 

When considering higher order collisional effects, the model and fit quality improved significantly. The same lines were again modelled, using this time a more advanced absorption model accounting for the speed-dependent pressure broadening effects, namely a reduction of the Hartmann-Tran profile (HTP)\cite{tennyson2014}. The limited signal-to-noise ratio of the remaining distortion, and the lack of pressure dependence study, made fitting the speed-dependent parameters directly from the data unreliable due to the correlation of the parameters. Instead, the HTP profile was fed with the pressure broadening coefficients and speed-dependent parameters measured in previous high precision studies \cite{kowzan2017, kowzan2020jqsrt} which were kept as fixed parameters. The fitting parameters were the gas sample concentration and temperature only, and the results agreed with the parameters retrieved from the Voigt profile fits within error bars. However, the residua of the HTP fits, shown in Fig.\ref{fig:HTP}(c) exhibit higher quality factors (calculated following the definition proposed by Cygan \textit{et al.}\cite{cygan2012}) and do not show any structure anymore.

\section{Conclusion}
In this study, we successfully developed a sensitive and precise cavity ring-down spectroscopy system capable of parallel acquisition over broad spectral ranges by coupling an optical frequency comb with time-resolved Fourier transform spectrometry. A key aspect of this approach was the decoupling of the frequency comb light from the Pound-Drever-Hall error signal, we were able to stabilize a high-finesse cavity while maintaining the frequency accuracy. This approach significantly enhances the achievable sensitivity of comb-based CRDS, allowing for high-precision gas absorption measurements. The retrieval of the absorption spectrum of CO broadened by Ar demonstrated the precision and accuracy of the technique, validating earlier findings obtained through other optical methods. Furthermore, the high quality factors achieved in absorption line retrieval enable the study of speed-dependent collisional effects on line profiles, with current limitations stemming from photodiode noise performance in detecting the interferograms. 

In comparison to recent advancements in comb-based CRDS as demonstrated by Liang \textit{et al.}\cite{liang2024}, our approach uses active stabilization of the cavity length to continuously match the comb spectrum. While this adds technical complexity to the experimental setup, it offers notable advantages, particularly in simplifying post-processing and achieving high-precision absorption measurements.  Additionally, the implementation of the locking loop facilitates synchronization of ring-down events with the optical path difference scan, which enhances the robustness of the technique and reduces sensitivity to small variations in the speed of the FTS translation stage. 

Unlike the dual-comb-based CRDS approache\cite{lisak2022}, where several interferograms were measured during a single cavity decay, our technique does not allow the absorption spectrum to be captured within a single cavity ring-down event. Instead, the decays must be repeated across the entire interferogram. However, this approach alleviates the requirement for an upper bandwidth limit high enough to resolve the interferograms, upper bandwidth now determined solely by the cavity parameters. As a result, this method yields an improved signal-to-noise ratio in interferogram acquisition, making it particularly advantageous for precision measurements.

The high sensitivity, broad spectral coverage, and precision of our setup make it a valuable tool for fundamental studies in molecular physics, such as collisional effects, as well as for various applications, from environmental monitoring and remote sensing to industrial processes control. The adaptability of comb-based FT-CRDS to other wavelength ranges, where traditional continuous-wave CRDS encounters limitations, will further expand its future potential for precision spectroscopic applications.

\begin{acknowledgments}
LR and RD acknowledge support from Agence Nationale de la Recherche (grant ANR-19-CE30-0038). PM acknowledges the support of National Science Centre, Poland project no. 2019/35/D/ST2/04114.
\end{acknowledgments}

\section*{Data Availability Statement}

The data that support the findings of this study are available from the corresponding author upon reasonable request.

\nocite{*}
\bibliography{biblio}

\begin{thebibliography}{33}%
\makeatletter
\providecommand \@ifxundefined [1]{%
 \@ifx{#1\undefined}
}%
\providecommand \@ifnum [1]{%
 \ifnum #1\expandafter \@firstoftwo
 \else \expandafter \@secondoftwo
 \fi
}%
\providecommand \@ifx [1]{%
 \ifx #1\expandafter \@firstoftwo
 \else \expandafter \@secondoftwo
 \fi
}%
\providecommand \natexlab [1]{#1}%
\providecommand \enquote  [1]{``#1''}%
\providecommand \bibnamefont  [1]{#1}%
\providecommand \bibfnamefont [1]{#1}%
\providecommand \citenamefont [1]{#1}%
\providecommand \href@noop [0]{\@secondoftwo}%
\providecommand \href [0]{\begingroup \@sanitize@url \@href}%
\providecommand \@href[1]{\@@startlink{#1}\@@href}%
\providecommand \@@href[1]{\endgroup#1\@@endlink}%
\providecommand \@sanitize@url [0]{\catcode `\\12\catcode `\$12\catcode `\&12\catcode `\#12\catcode `\^12\catcode `\_12\catcode `\%12\relax}%
\providecommand \@@startlink[1]{}%
\providecommand \@@endlink[0]{}%
\providecommand \url  [0]{\begingroup\@sanitize@url \@url }%
\providecommand \@url [1]{\endgroup\@href {#1}{\urlprefix }}%
\providecommand \urlprefix  [0]{URL }%
\providecommand \Eprint [0]{\href }%
\providecommand \doibase [0]{http://dx.doi.org/}%
\providecommand \selectlanguage [0]{\@gobble}%
\providecommand \bibinfo  [0]{\@secondoftwo}%
\providecommand \bibfield  [0]{\@secondoftwo}%
\providecommand \translation [1]{[#1]}%
\providecommand \BibitemOpen [0]{}%
\providecommand \bibitemStop [0]{}%
\providecommand \bibitemNoStop [0]{.\EOS\space}%
\providecommand \EOS [0]{\spacefactor3000\relax}%
\providecommand \BibitemShut  [1]{\csname bibitem#1\endcsname}%
\let\auto@bib@innerbib\@empty
\bibitem [{\citenamefont {Long}\ \emph {et~al.}(2014)\citenamefont {Long}, \citenamefont {Truong}, \citenamefont {van Zee}, \citenamefont {Plusquellic},\ and\ \citenamefont {Hodges}}]{long2014}%
  \BibitemOpen
  \bibfield  {author} {\bibinfo {author} {\bibfnamefont {D.~A.}\ \bibnamefont {Long}}, \bibinfo {author} {\bibfnamefont {G.-W.}\ \bibnamefont {Truong}}, \bibinfo {author} {\bibfnamefont {R.~D.}\ \bibnamefont {van Zee}}, \bibinfo {author} {\bibfnamefont {D.~F.}\ \bibnamefont {Plusquellic}}, \ and\ \bibinfo {author} {\bibfnamefont {J.~T.}\ \bibnamefont {Hodges}},\ }\bibfield  {title} {\enquote {\bibinfo {title} {Frequency-agile, rapid scanning spectroscopy: absorption sensitivity of $2\times 10^{-12}$ cm$^{-1}$ hz$^{-1/2}$ with a tunable diode laser},}\ }\href@noop {} {\bibfield  {journal} {\bibinfo  {journal} {Applied Physics B}\ }\textbf {\bibinfo {volume} {114}},\ \bibinfo {pages} {489--495} (\bibinfo {year} {2014})}\BibitemShut {NoStop}%
\bibitem [{\citenamefont {Long}\ \emph {et~al.}(2015)\citenamefont {Long}, \citenamefont {Wojtewicz}, \citenamefont {Miller},\ and\ \citenamefont {Hodges}}]{long2015}%
  \BibitemOpen
  \bibfield  {author} {\bibinfo {author} {\bibfnamefont {D.~A.}\ \bibnamefont {Long}}, \bibinfo {author} {\bibfnamefont {S.}~\bibnamefont {Wojtewicz}}, \bibinfo {author} {\bibfnamefont {C.~E.}\ \bibnamefont {Miller}}, \ and\ \bibinfo {author} {\bibfnamefont {J.~T.}\ \bibnamefont {Hodges}},\ }\bibfield  {title} {\enquote {\bibinfo {title} {Frequency-agile, rapid scanning cavity ring-down spectroscopy (fars-crds) measurements of the (30012)<-(00001) near-infrared carbon dioxide band},}\ }\href@noop {} {\bibfield  {journal} {\bibinfo  {journal} {Journal of Quantitative Spectroscopy and Radiative Transfer}\ }\textbf {\bibinfo {volume} {161}},\ \bibinfo {pages} {35--40} (\bibinfo {year} {2015})}\BibitemShut {NoStop}%
\bibitem [{\citenamefont {Burkart}\ \emph {et~al.}(2015)\citenamefont {Burkart}, \citenamefont {Sala}, \citenamefont {Romanini}, \citenamefont {Marangoni}, \citenamefont {Campargue},\ and\ \citenamefont {Kassi}}]{burkart2015}%
  \BibitemOpen
  \bibfield  {author} {\bibinfo {author} {\bibfnamefont {J.}~\bibnamefont {Burkart}}, \bibinfo {author} {\bibfnamefont {T.}~\bibnamefont {Sala}}, \bibinfo {author} {\bibfnamefont {D.}~\bibnamefont {Romanini}}, \bibinfo {author} {\bibfnamefont {M.}~\bibnamefont {Marangoni}}, \bibinfo {author} {\bibfnamefont {A.}~\bibnamefont {Campargue}}, \ and\ \bibinfo {author} {\bibfnamefont {S.}~\bibnamefont {Kassi}},\ }\bibfield  {title} {\enquote {\bibinfo {title} {Communication: Saturated co$_2$ absorption near 1.6 $\mu$m for kilohertz-accuracy transition frequencies},}\ }\href@noop {} {\bibfield  {journal} {\bibinfo  {journal} {The Journal of chemical physics}\ }\textbf {\bibinfo {volume} {142}} (\bibinfo {year} {2015})}\BibitemShut {NoStop}%
\bibitem [{\citenamefont {Votava}\ \emph {et~al.}(2022)\citenamefont {Votava}, \citenamefont {Kassi}, \citenamefont {Campargue},\ and\ \citenamefont {Romanini}}]{votava2022}%
  \BibitemOpen
  \bibfield  {author} {\bibinfo {author} {\bibfnamefont {O.}~\bibnamefont {Votava}}, \bibinfo {author} {\bibfnamefont {S.}~\bibnamefont {Kassi}}, \bibinfo {author} {\bibfnamefont {A.}~\bibnamefont {Campargue}}, \ and\ \bibinfo {author} {\bibfnamefont {D.}~\bibnamefont {Romanini}},\ }\bibfield  {title} {\enquote {\bibinfo {title} {Comb coherence-transfer and cavity ring-down saturation spectroscopy around 1.65 $\mu$m: khz-accurate frequencies of transitions in the 2$\nu$3 band of $^12$ch$_4$},}\ }\href {\doibase 10.1039/D1CP04989E} {\bibfield  {journal} {\bibinfo  {journal} {Phys. Chem. Chem. Phys.}\ }\textbf {\bibinfo {volume} {24}},\ \bibinfo {pages} {4157--4173} (\bibinfo {year} {2022})}\BibitemShut {NoStop}%
\bibitem [{\citenamefont {Gotti}\ \emph {et~al.}(2020)\citenamefont {Gotti}, \citenamefont {Puppe}, \citenamefont {Mayzlin}, \citenamefont {Robinson-Tait}, \citenamefont {W{\'o}jtewicz}, \citenamefont {Gatti}, \citenamefont {Alsaif}, \citenamefont {Lamperti}, \citenamefont {Laporta}, \citenamefont {Rohde} \emph {et~al.}}]{gotti2020}%
  \BibitemOpen
  \bibfield  {author} {\bibinfo {author} {\bibfnamefont {R.}~\bibnamefont {Gotti}}, \bibinfo {author} {\bibfnamefont {T.}~\bibnamefont {Puppe}}, \bibinfo {author} {\bibfnamefont {Y.}~\bibnamefont {Mayzlin}}, \bibinfo {author} {\bibfnamefont {J.}~\bibnamefont {Robinson-Tait}}, \bibinfo {author} {\bibfnamefont {S.}~\bibnamefont {W{\'o}jtewicz}}, \bibinfo {author} {\bibfnamefont {D.}~\bibnamefont {Gatti}}, \bibinfo {author} {\bibfnamefont {B.}~\bibnamefont {Alsaif}}, \bibinfo {author} {\bibfnamefont {M.}~\bibnamefont {Lamperti}}, \bibinfo {author} {\bibfnamefont {P.}~\bibnamefont {Laporta}}, \bibinfo {author} {\bibfnamefont {F.}~\bibnamefont {Rohde}},  \emph {et~al.},\ }\bibfield  {title} {\enquote {\bibinfo {title} {Comb-locked frequency-swept synthesizer for high precision broadband spectroscopy},}\ }\href@noop {} {\bibfield  {journal} {\bibinfo  {journal} {Scientific Reports}\ }\textbf {\bibinfo {volume} {10}},\ \bibinfo {pages} {2523} (\bibinfo {year} {2020})}\BibitemShut {NoStop}%
\bibitem [{\citenamefont {Engeln}\ and\ \citenamefont {Meijer}(1996)}]{engeln1996}%
  \BibitemOpen
  \bibfield  {author} {\bibinfo {author} {\bibfnamefont {R.}~\bibnamefont {Engeln}}\ and\ \bibinfo {author} {\bibfnamefont {G.}~\bibnamefont {Meijer}},\ }\bibfield  {title} {\enquote {\bibinfo {title} {{A Fourier transform cavity ring down spectrometer}},}\ }\href {\doibase 10.1063/1.1147092} {\bibfield  {journal} {\bibinfo  {journal} {Review of Scientific Instruments}\ }\textbf {\bibinfo {volume} {67}},\ \bibinfo {pages} {2708--2713} (\bibinfo {year} {1996})},\ \Eprint {http://arxiv.org/abs/https://pubs.aip.org/aip/rsi/article-pdf/67/8/2708/19172654/2708\_1\_online.pdf} {https://pubs.aip.org/aip/rsi/article-pdf/67/8/2708/19172654/2708\_1\_online.pdf} \BibitemShut {NoStop}%
\bibitem [{\citenamefont {Scherer}(1998)}]{scherer1998}%
  \BibitemOpen
  \bibfield  {author} {\bibinfo {author} {\bibfnamefont {J.~J.}\ \bibnamefont {Scherer}},\ }\bibfield  {title} {\enquote {\bibinfo {title} {Ringdown spectral photography},}\ }\href@noop {} {\bibfield  {journal} {\bibinfo  {journal} {Chemical Physics Letters}\ }\textbf {\bibinfo {volume} {292}},\ \bibinfo {pages} {143--153} (\bibinfo {year} {1998})}\BibitemShut {NoStop}%
\bibitem [{\citenamefont {Czy{\.z}ewski}\ \emph {et~al.}(2001)\citenamefont {Czy{\.z}ewski}, \citenamefont {Chudzy{\'n}ski}, \citenamefont {Ernst}, \citenamefont {Karasi{\'n}ski}, \citenamefont {Pietruczuk}, \citenamefont {Skubiszak}, \citenamefont {Stacewicz}, \citenamefont {Stelmaszczyk}, \citenamefont {Koch}, \citenamefont {Rairoux} \emph {et~al.}}]{czyzewski2001}%
  \BibitemOpen
  \bibfield  {author} {\bibinfo {author} {\bibfnamefont {A.}~\bibnamefont {Czy{\.z}ewski}}, \bibinfo {author} {\bibfnamefont {S.}~\bibnamefont {Chudzy{\'n}ski}}, \bibinfo {author} {\bibfnamefont {K.}~\bibnamefont {Ernst}}, \bibinfo {author} {\bibfnamefont {G.}~\bibnamefont {Karasi{\'n}ski}}, \bibinfo {author} {\bibfnamefont {A.}~\bibnamefont {Pietruczuk}}, \bibinfo {author} {\bibfnamefont {W.}~\bibnamefont {Skubiszak}}, \bibinfo {author} {\bibfnamefont {T.}~\bibnamefont {Stacewicz}}, \bibinfo {author} {\bibfnamefont {K.}~\bibnamefont {Stelmaszczyk}}, \bibinfo {author} {\bibfnamefont {B.}~\bibnamefont {Koch}}, \bibinfo {author} {\bibfnamefont {P.}~\bibnamefont {Rairoux}},  \emph {et~al.},\ }\bibfield  {title} {\enquote {\bibinfo {title} {Cavity ring-down spectrography},}\ }\href@noop {} {\bibfield  {journal} {\bibinfo  {journal} {Optics communications}\ }\textbf {\bibinfo {volume} {191}},\ \bibinfo {pages} {271--275} (\bibinfo {year} {2001})}\BibitemShut {NoStop}%
\bibitem [{\citenamefont {Thorpe}\ \emph {et~al.}(2006)\citenamefont {Thorpe}, \citenamefont {Moll}, \citenamefont {Jones}, \citenamefont {Safdi},\ and\ \citenamefont {Ye}}]{thorpe2006}%
  \BibitemOpen
  \bibfield  {author} {\bibinfo {author} {\bibfnamefont {M.~J.}\ \bibnamefont {Thorpe}}, \bibinfo {author} {\bibfnamefont {K.~D.}\ \bibnamefont {Moll}}, \bibinfo {author} {\bibfnamefont {R.~J.}\ \bibnamefont {Jones}}, \bibinfo {author} {\bibfnamefont {B.}~\bibnamefont {Safdi}}, \ and\ \bibinfo {author} {\bibfnamefont {J.}~\bibnamefont {Ye}},\ }\bibfield  {title} {\enquote {\bibinfo {title} {Broadband cavity ringdown spectroscopy for sensitive and rapid molecular detection},}\ }\href {\doibase 10.1126/science.1123921} {\bibfield  {journal} {\bibinfo  {journal} {Science}\ }\textbf {\bibinfo {volume} {311}},\ \bibinfo {pages} {1595--1599} (\bibinfo {year} {2006})},\ \Eprint {http://arxiv.org/abs/https://www.science.org/doi/pdf/10.1126/science.1123921} {https://www.science.org/doi/pdf/10.1126/science.1123921} \BibitemShut {NoStop}%
\bibitem [{\citenamefont {Lisak}\ \emph {et~al.}(2022)\citenamefont {Lisak}, \citenamefont {Charczun}, \citenamefont {Nishiyama}, \citenamefont {Voumard}, \citenamefont {Wildi}, \citenamefont {Kowzan}, \citenamefont {Brasch}, \citenamefont {Herr}, \citenamefont {Fleisher}, \citenamefont {Hodges} \emph {et~al.}}]{lisak2022}%
  \BibitemOpen
  \bibfield  {author} {\bibinfo {author} {\bibfnamefont {D.}~\bibnamefont {Lisak}}, \bibinfo {author} {\bibfnamefont {D.}~\bibnamefont {Charczun}}, \bibinfo {author} {\bibfnamefont {A.}~\bibnamefont {Nishiyama}}, \bibinfo {author} {\bibfnamefont {T.}~\bibnamefont {Voumard}}, \bibinfo {author} {\bibfnamefont {T.}~\bibnamefont {Wildi}}, \bibinfo {author} {\bibfnamefont {G.}~\bibnamefont {Kowzan}}, \bibinfo {author} {\bibfnamefont {V.}~\bibnamefont {Brasch}}, \bibinfo {author} {\bibfnamefont {T.}~\bibnamefont {Herr}}, \bibinfo {author} {\bibfnamefont {A.~J.}\ \bibnamefont {Fleisher}}, \bibinfo {author} {\bibfnamefont {J.~T.}\ \bibnamefont {Hodges}},  \emph {et~al.},\ }\bibfield  {title} {\enquote {\bibinfo {title} {Dual-comb cavity ring-down spectroscopy},}\ }\href@noop {} {\bibfield  {journal} {\bibinfo  {journal} {Scientific reports}\ }\textbf {\bibinfo {volume} {12}},\ \bibinfo {pages} {2377} (\bibinfo {year} {2022})}\BibitemShut {NoStop}%
\bibitem [{\citenamefont {Dubroeucq}\ and\ \citenamefont {Rutkowski}(2022)}]{dubroeucq2022}%
  \BibitemOpen
  \bibfield  {author} {\bibinfo {author} {\bibfnamefont {R.}~\bibnamefont {Dubroeucq}}\ and\ \bibinfo {author} {\bibfnamefont {L.}~\bibnamefont {Rutkowski}},\ }\bibfield  {title} {\enquote {\bibinfo {title} {Optical frequency comb fourier transform cavity ring-down spectroscopy},}\ }\href@noop {} {\bibfield  {journal} {\bibinfo  {journal} {Optics Express}\ }\textbf {\bibinfo {volume} {30}},\ \bibinfo {pages} {13594--13602} (\bibinfo {year} {2022})}\BibitemShut {NoStop}%
\bibitem [{\citenamefont {Black}(2001)}]{black2001}%
  \BibitemOpen
  \bibfield  {author} {\bibinfo {author} {\bibfnamefont {E.~D.}\ \bibnamefont {Black}},\ }\bibfield  {title} {\enquote {\bibinfo {title} {An introduction to pound-drever-hall laser frequency stabilization},}\ }\href@noop {} {\bibfield  {journal} {\bibinfo  {journal} {American journal of physics}\ }\textbf {\bibinfo {volume} {69}},\ \bibinfo {pages} {79--87} (\bibinfo {year} {2001})}\BibitemShut {NoStop}%
\bibitem [{\citenamefont {Liang}\ \emph {et~al.}(2024)\citenamefont {Liang}, \citenamefont {Bisht}, \citenamefont {Scheck}, \citenamefont {Schunemann},\ and\ \citenamefont {Ye}}]{liang2024}%
  \BibitemOpen
  \bibfield  {author} {\bibinfo {author} {\bibfnamefont {Q.}~\bibnamefont {Liang}}, \bibinfo {author} {\bibfnamefont {A.}~\bibnamefont {Bisht}}, \bibinfo {author} {\bibfnamefont {A.}~\bibnamefont {Scheck}}, \bibinfo {author} {\bibfnamefont {P.~G.}\ \bibnamefont {Schunemann}}, \ and\ \bibinfo {author} {\bibfnamefont {J.}~\bibnamefont {Ye}},\ }\bibfield  {title} {\enquote {\bibinfo {title} {Modulated ringdown comb interferometry for next-generation high complexity trace gas sensing},}\ }\href@noop {} {\bibfield  {journal} {\bibinfo  {journal} {arXiv preprint arXiv:2406.03609}\ } (\bibinfo {year} {2024})}\BibitemShut {NoStop}%
\bibitem [{\citenamefont {Gerwert}(1999)}]{gerwert1999}%
  \BibitemOpen
  \bibfield  {author} {\bibinfo {author} {\bibfnamefont {K.}~\bibnamefont {Gerwert}},\ }\bibfield  {title} {\enquote {\bibinfo {title} {Molecular reaction mechanisms of proteins monitored by time-resolved ftir-spectroscopy},}\ }\href@noop {} {\  (\bibinfo {year} {1999})}\BibitemShut {NoStop}%
\bibitem [{\citenamefont {Picqu{\'e}}\ and\ \citenamefont {Guelachvili}(2000)}]{picque2000}%
  \BibitemOpen
  \bibfield  {author} {\bibinfo {author} {\bibfnamefont {N.}~\bibnamefont {Picqu{\'e}}}\ and\ \bibinfo {author} {\bibfnamefont {G.}~\bibnamefont {Guelachvili}},\ }\bibfield  {title} {\enquote {\bibinfo {title} {High-information time-resolved fourier transform spectroscopy at work},}\ }\href@noop {} {\bibfield  {journal} {\bibinfo  {journal} {Applied optics}\ }\textbf {\bibinfo {volume} {39}},\ \bibinfo {pages} {3984--3990} (\bibinfo {year} {2000})}\BibitemShut {NoStop}%
\bibitem [{\citenamefont {Griffiths}\ and\ \citenamefont {James}(2007)}]{griffiths2007}%
  \BibitemOpen
  \bibfield  {author} {\bibinfo {author} {\bibfnamefont {P.~R.}\ \bibnamefont {Griffiths}}\ and\ \bibinfo {author} {\bibfnamefont {A.}~\bibnamefont {James}},\ }\bibfield  {title} {\enquote {\bibinfo {title} {De haseth, fourier transform infrared spectrometry},}\ }\href@noop {} {\bibfield  {journal} {\bibinfo  {journal} {John Wiley \& Sons World Academy of Science. J Eng Technol}\ }\textbf {\bibinfo {volume} {5}},\ \bibinfo {pages} {1473--1478} (\bibinfo {year} {2007})}\BibitemShut {NoStop}%
\bibitem [{\citenamefont {Foltynowicz}\ \emph {et~al.}(2011)\citenamefont {Foltynowicz}, \citenamefont {Ban}, \citenamefont {Mas{\l}owski}, \citenamefont {Adler},\ and\ \citenamefont {Ye}}]{foltynowicz2011}%
  \BibitemOpen
  \bibfield  {author} {\bibinfo {author} {\bibfnamefont {A.}~\bibnamefont {Foltynowicz}}, \bibinfo {author} {\bibfnamefont {T.}~\bibnamefont {Ban}}, \bibinfo {author} {\bibfnamefont {P.}~\bibnamefont {Mas{\l}owski}}, \bibinfo {author} {\bibfnamefont {F.}~\bibnamefont {Adler}}, \ and\ \bibinfo {author} {\bibfnamefont {J.}~\bibnamefont {Ye}},\ }\bibfield  {title} {\enquote {\bibinfo {title} {Quantum-noise-limited optical frequency comb spectroscopy},}\ }\href@noop {} {\bibfield  {journal} {\bibinfo  {journal} {Physical review letters}\ }\textbf {\bibinfo {volume} {107}},\ \bibinfo {pages} {233002} (\bibinfo {year} {2011})}\BibitemShut {NoStop}%
\bibitem [{\citenamefont {Nishiyama}\ \emph {et~al.}(2020)\citenamefont {Nishiyama}, \citenamefont {Kowzan}, \citenamefont {Charczun}, \citenamefont {Trawi{\'n}ski},\ and\ \citenamefont {Mas{\l}owski}}]{nishiyama2020}%
  \BibitemOpen
  \bibfield  {author} {\bibinfo {author} {\bibfnamefont {A.}~\bibnamefont {Nishiyama}}, \bibinfo {author} {\bibfnamefont {G.}~\bibnamefont {Kowzan}}, \bibinfo {author} {\bibfnamefont {D.}~\bibnamefont {Charczun}}, \bibinfo {author} {\bibfnamefont {R.~S.}\ \bibnamefont {Trawi{\'n}ski}}, \ and\ \bibinfo {author} {\bibfnamefont {P.}~\bibnamefont {Mas{\l}owski}},\ }\bibfield  {title} {\enquote {\bibinfo {title} {Optical frequency comb-based cavity-enhanced fourier-transform spectroscopy: Application to collisional line-shape study},}\ }\href@noop {} {\bibfield  {journal} {\bibinfo  {journal} {Chinese Journal of Chemical Physics}\ }\textbf {\bibinfo {volume} {33}},\ \bibinfo {pages} {23--30} (\bibinfo {year} {2020})}\BibitemShut {NoStop}%
\bibitem [{\citenamefont {{\'S}liwczy{\'n}ski}\ \emph {et~al.}(2013)\citenamefont {{\'S}liwczy{\'n}ski}, \citenamefont {Krehlik}, \citenamefont {Czubla}, \citenamefont {Buczek},\ and\ \citenamefont {Lipi{\'n}ski}}]{sliwczynski2013}%
  \BibitemOpen
  \bibfield  {author} {\bibinfo {author} {\bibfnamefont {{\L}.}~\bibnamefont {{\'S}liwczy{\'n}ski}}, \bibinfo {author} {\bibfnamefont {P.}~\bibnamefont {Krehlik}}, \bibinfo {author} {\bibfnamefont {A.}~\bibnamefont {Czubla}}, \bibinfo {author} {\bibfnamefont {{\L}.}~\bibnamefont {Buczek}}, \ and\ \bibinfo {author} {\bibfnamefont {M.}~\bibnamefont {Lipi{\'n}ski}},\ }\bibfield  {title} {\enquote {\bibinfo {title} {Dissemination of time and rf frequency via a stabilized fibre optic link over a distance of 420 km},}\ }\href@noop {} {\bibfield  {journal} {\bibinfo  {journal} {Metrologia}\ }\textbf {\bibinfo {volume} {50}},\ \bibinfo {pages} {133} (\bibinfo {year} {2013})}\BibitemShut {NoStop}%
\bibitem [{\citenamefont {Morzy{\'n}ski}\ \emph {et~al.}(2015)\citenamefont {Morzy{\'n}ski}, \citenamefont {Bober}, \citenamefont {Bartoszek-Bober}, \citenamefont {Nawrocki}, \citenamefont {Krehlik}, \citenamefont {{\'S}liwczy{\'n}ski}, \citenamefont {Lipi{\'n}ski}, \citenamefont {Mas{\l}owski}, \citenamefont {Cygan}, \citenamefont {Dunst} \emph {et~al.}}]{morzynski2015}%
  \BibitemOpen
  \bibfield  {author} {\bibinfo {author} {\bibfnamefont {P.}~\bibnamefont {Morzy{\'n}ski}}, \bibinfo {author} {\bibfnamefont {M.}~\bibnamefont {Bober}}, \bibinfo {author} {\bibfnamefont {D.}~\bibnamefont {Bartoszek-Bober}}, \bibinfo {author} {\bibfnamefont {J.}~\bibnamefont {Nawrocki}}, \bibinfo {author} {\bibfnamefont {P.}~\bibnamefont {Krehlik}}, \bibinfo {author} {\bibfnamefont {{\L}.}~\bibnamefont {{\'S}liwczy{\'n}ski}}, \bibinfo {author} {\bibfnamefont {M.}~\bibnamefont {Lipi{\'n}ski}}, \bibinfo {author} {\bibfnamefont {P.}~\bibnamefont {Mas{\l}owski}}, \bibinfo {author} {\bibfnamefont {A.}~\bibnamefont {Cygan}}, \bibinfo {author} {\bibfnamefont {P.}~\bibnamefont {Dunst}},  \emph {et~al.},\ }\bibfield  {title} {\enquote {\bibinfo {title} {Absolute measurement of the 1s0-3p0 clock transition in neutral 88sr over the 330 km-long stabilized fibre optic link},}\ }\href@noop {} {\bibfield  {journal} {\bibinfo  {journal} {Scientific Reports}\ }\textbf {\bibinfo {volume} {5}},\ \bibinfo {pages} {17495}
  (\bibinfo {year} {2015})}\BibitemShut {NoStop}%
\bibitem [{\citenamefont {Kowzan}\ \emph {et~al.}(2017)\citenamefont {Kowzan}, \citenamefont {Stec}, \citenamefont {Zaborowski}, \citenamefont {W{\'o}jtewicz}, \citenamefont {Cygan}, \citenamefont {Lisak}, \citenamefont {Mas{\l}owski},\ and\ \citenamefont {Trawi{\'n}ski}}]{kowzan2017}%
  \BibitemOpen
  \bibfield  {author} {\bibinfo {author} {\bibfnamefont {G.}~\bibnamefont {Kowzan}}, \bibinfo {author} {\bibfnamefont {K.}~\bibnamefont {Stec}}, \bibinfo {author} {\bibfnamefont {M.}~\bibnamefont {Zaborowski}}, \bibinfo {author} {\bibfnamefont {S.}~\bibnamefont {W{\'o}jtewicz}}, \bibinfo {author} {\bibfnamefont {A.}~\bibnamefont {Cygan}}, \bibinfo {author} {\bibfnamefont {D.}~\bibnamefont {Lisak}}, \bibinfo {author} {\bibfnamefont {P.}~\bibnamefont {Mas{\l}owski}}, \ and\ \bibinfo {author} {\bibfnamefont {R.}~\bibnamefont {Trawi{\'n}ski}},\ }\bibfield  {title} {\enquote {\bibinfo {title} {Line positions, pressure broadening and shift coefficients for the second overtone transitions of carbon monoxide in argon},}\ }\href@noop {} {\bibfield  {journal} {\bibinfo  {journal} {Journal of Quantitative Spectroscopy and Radiative Transfer}\ }\textbf {\bibinfo {volume} {191}},\ \bibinfo {pages} {46--54} (\bibinfo {year} {2017})}\BibitemShut {NoStop}%
\bibitem [{\citenamefont {Kowzan}\ \emph {et~al.}(2020{\natexlab{a}})\citenamefont {Kowzan}, \citenamefont {Wcis{\l}o}, \citenamefont {S{\l}owi{\'n}ski}, \citenamefont {Mas{\l}owski}, \citenamefont {Viel},\ and\ \citenamefont {Thibault}}]{kowzan2020jqsrt}%
  \BibitemOpen
  \bibfield  {author} {\bibinfo {author} {\bibfnamefont {G.}~\bibnamefont {Kowzan}}, \bibinfo {author} {\bibfnamefont {P.}~\bibnamefont {Wcis{\l}o}}, \bibinfo {author} {\bibfnamefont {M.}~\bibnamefont {S{\l}owi{\'n}ski}}, \bibinfo {author} {\bibfnamefont {P.}~\bibnamefont {Mas{\l}owski}}, \bibinfo {author} {\bibfnamefont {A.}~\bibnamefont {Viel}}, \ and\ \bibinfo {author} {\bibfnamefont {F.}~\bibnamefont {Thibault}},\ }\bibfield  {title} {\enquote {\bibinfo {title} {Fully quantum calculations of the line-shape parameters for the hartmann-tran profile: A co-ar case study},}\ }\href@noop {} {\bibfield  {journal} {\bibinfo  {journal} {Journal of Quantitative Spectroscopy and Radiative Transfer}\ }\textbf {\bibinfo {volume} {243}},\ \bibinfo {pages} {106803} (\bibinfo {year} {2020}{\natexlab{a}})}\BibitemShut {NoStop}%
\bibitem [{\citenamefont {Kowzan}\ \emph {et~al.}(2020{\natexlab{b}})\citenamefont {Kowzan}, \citenamefont {Cybulski}, \citenamefont {Wcis{\l}o}, \citenamefont {S{\l}owi{\'n}ski}, \citenamefont {Viel}, \citenamefont {Mas{\l}owski},\ and\ \citenamefont {Thibault}}]{kowzan2020pra}%
  \BibitemOpen
  \bibfield  {author} {\bibinfo {author} {\bibfnamefont {G.}~\bibnamefont {Kowzan}}, \bibinfo {author} {\bibfnamefont {H.}~\bibnamefont {Cybulski}}, \bibinfo {author} {\bibfnamefont {P.}~\bibnamefont {Wcis{\l}o}}, \bibinfo {author} {\bibfnamefont {M.}~\bibnamefont {S{\l}owi{\'n}ski}}, \bibinfo {author} {\bibfnamefont {A.}~\bibnamefont {Viel}}, \bibinfo {author} {\bibfnamefont {P.}~\bibnamefont {Mas{\l}owski}}, \ and\ \bibinfo {author} {\bibfnamefont {F.}~\bibnamefont {Thibault}},\ }\bibfield  {title} {\enquote {\bibinfo {title} {Subpercent agreement between ab initio and experimental collision-induced line shapes of carbon monoxide perturbed by argon},}\ }\href@noop {} {\bibfield  {journal} {\bibinfo  {journal} {Physical Review A}\ }\textbf {\bibinfo {volume} {102}},\ \bibinfo {pages} {012821} (\bibinfo {year} {2020}{\natexlab{b}})}\BibitemShut {NoStop}%
\bibitem [{\citenamefont {Reed}\ \emph {et~al.}(2023)\citenamefont {Reed}, \citenamefont {Tran}, \citenamefont {Ngo}, \citenamefont {Hartmann},\ and\ \citenamefont {Hodges}}]{reed2023}%
  \BibitemOpen
  \bibfield  {author} {\bibinfo {author} {\bibfnamefont {Z.~D.}\ \bibnamefont {Reed}}, \bibinfo {author} {\bibfnamefont {H.}~\bibnamefont {Tran}}, \bibinfo {author} {\bibfnamefont {H.~N.}\ \bibnamefont {Ngo}}, \bibinfo {author} {\bibfnamefont {J.-M.}\ \bibnamefont {Hartmann}}, \ and\ \bibinfo {author} {\bibfnamefont {J.~T.}\ \bibnamefont {Hodges}},\ }\bibfield  {title} {\enquote {\bibinfo {title} {Effect of non-markovian collisions on measured integrated line shapes of co},}\ }\href@noop {} {\bibfield  {journal} {\bibinfo  {journal} {Physical Review Letters}\ }\textbf {\bibinfo {volume} {130}},\ \bibinfo {pages} {143001} (\bibinfo {year} {2023})}\BibitemShut {NoStop}%
\bibitem [{\citenamefont {Gordon}\ \emph {et~al.}(2022)\citenamefont {Gordon}, \citenamefont {Rothman}, \citenamefont {Hargreaves}, \citenamefont {Hashemi}, \citenamefont {Karlovets}, \citenamefont {Skinner}, \citenamefont {Conway}, \citenamefont {Hill}, \citenamefont {Kochanov}, \citenamefont {Tan} \emph {et~al.}}]{gordon2022}%
  \BibitemOpen
  \bibfield  {author} {\bibinfo {author} {\bibfnamefont {I.~E.}\ \bibnamefont {Gordon}}, \bibinfo {author} {\bibfnamefont {L.~S.}\ \bibnamefont {Rothman}}, \bibinfo {author} {\bibfnamefont {R.}~\bibnamefont {Hargreaves}}, \bibinfo {author} {\bibfnamefont {R.}~\bibnamefont {Hashemi}}, \bibinfo {author} {\bibfnamefont {E.~V.}\ \bibnamefont {Karlovets}}, \bibinfo {author} {\bibfnamefont {F.}~\bibnamefont {Skinner}}, \bibinfo {author} {\bibfnamefont {E.~K.}\ \bibnamefont {Conway}}, \bibinfo {author} {\bibfnamefont {C.}~\bibnamefont {Hill}}, \bibinfo {author} {\bibfnamefont {R.~V.}\ \bibnamefont {Kochanov}}, \bibinfo {author} {\bibfnamefont {Y.}~\bibnamefont {Tan}},  \emph {et~al.},\ }\bibfield  {title} {\enquote {\bibinfo {title} {The hitran2020 molecular spectroscopic database},}\ }\href@noop {} {\bibfield  {journal} {\bibinfo  {journal} {Journal of quantitative spectroscopy and radiative transfer}\ }\textbf {\bibinfo {volume} {277}},\ \bibinfo {pages} {107949} (\bibinfo {year} {2022})}\BibitemShut {NoStop}%
\bibitem [{\citenamefont {Bouanich}\ and\ \citenamefont {Haeusler}(1972)}]{bouanich1972}%
  \BibitemOpen
  \bibfield  {author} {\bibinfo {author} {\bibfnamefont {J.-P.}\ \bibnamefont {Bouanich}}\ and\ \bibinfo {author} {\bibfnamefont {C.}~\bibnamefont {Haeusler}},\ }\bibfield  {title} {\enquote {\bibinfo {title} {Linewidths of carbon monoxide self-broadening and broadened by argon and nitrogen},}\ }\href@noop {} {\bibfield  {journal} {\bibinfo  {journal} {Journal of Quantitative Spectroscopy and Radiative Transfer}\ }\textbf {\bibinfo {volume} {12}},\ \bibinfo {pages} {695--702} (\bibinfo {year} {1972})}\BibitemShut {NoStop}%
\bibitem [{\citenamefont {Cygan}\ \emph {et~al.}(2012)\citenamefont {Cygan}, \citenamefont {Lisak}, \citenamefont {W{\'o}jtewicz}, \citenamefont {Domys{\l}awska}, \citenamefont {Hodges}, \citenamefont {Trawi{\'n}ski},\ and\ \citenamefont {Ciury{\l}o}}]{cygan2012}%
  \BibitemOpen
  \bibfield  {author} {\bibinfo {author} {\bibfnamefont {A.}~\bibnamefont {Cygan}}, \bibinfo {author} {\bibfnamefont {D.}~\bibnamefont {Lisak}}, \bibinfo {author} {\bibfnamefont {S.}~\bibnamefont {W{\'o}jtewicz}}, \bibinfo {author} {\bibfnamefont {J.}~\bibnamefont {Domys{\l}awska}}, \bibinfo {author} {\bibfnamefont {J.~T.}\ \bibnamefont {Hodges}}, \bibinfo {author} {\bibfnamefont {R.}~\bibnamefont {Trawi{\'n}ski}}, \ and\ \bibinfo {author} {\bibfnamefont {R.}~\bibnamefont {Ciury{\l}o}},\ }\bibfield  {title} {\enquote {\bibinfo {title} {High-signal-to-noise-ratio laser technique for accurate measurements of spectral line parameters},}\ }\href@noop {} {\bibfield  {journal} {\bibinfo  {journal} {Physical Review A—Atomic, Molecular, and Optical Physics}\ }\textbf {\bibinfo {volume} {85}},\ \bibinfo {pages} {022508} (\bibinfo {year} {2012})}\BibitemShut {NoStop}%
\bibitem [{\citenamefont {Tennyson}\ \emph {et~al.}(2014)\citenamefont {Tennyson}, \citenamefont {Bernath}, \citenamefont {Campargue}, \citenamefont {Cs{\'a}sz{\'a}r}, \citenamefont {Daumont}, \citenamefont {Gamache}, \citenamefont {Hodges}, \citenamefont {Lisak}, \citenamefont {Naumenko}, \citenamefont {Rothman} \emph {et~al.}}]{tennyson2014}%
  \BibitemOpen
  \bibfield  {author} {\bibinfo {author} {\bibfnamefont {J.}~\bibnamefont {Tennyson}}, \bibinfo {author} {\bibfnamefont {P.~F.}\ \bibnamefont {Bernath}}, \bibinfo {author} {\bibfnamefont {A.}~\bibnamefont {Campargue}}, \bibinfo {author} {\bibfnamefont {A.~G.}\ \bibnamefont {Cs{\'a}sz{\'a}r}}, \bibinfo {author} {\bibfnamefont {L.}~\bibnamefont {Daumont}}, \bibinfo {author} {\bibfnamefont {R.~R.}\ \bibnamefont {Gamache}}, \bibinfo {author} {\bibfnamefont {J.~T.}\ \bibnamefont {Hodges}}, \bibinfo {author} {\bibfnamefont {D.}~\bibnamefont {Lisak}}, \bibinfo {author} {\bibfnamefont {O.~V.}\ \bibnamefont {Naumenko}}, \bibinfo {author} {\bibfnamefont {L.~S.}\ \bibnamefont {Rothman}},  \emph {et~al.},\ }\bibfield  {title} {\enquote {\bibinfo {title} {Recommended isolated-line profile for representing high-resolution spectroscopic transitions (iupac technical report)},}\ }\href@noop {} {\bibfield  {journal} {\bibinfo  {journal} {Pure and Applied Chemistry}\ }\textbf {\bibinfo {volume} {86}},\ \bibinfo {pages}
  {1931--1943} (\bibinfo {year} {2014})}\BibitemShut {NoStop}%
\bibitem [{\citenamefont {Ngo}\ \emph {et~al.}(2017)\citenamefont {Ngo}, \citenamefont {Lin}, \citenamefont {Hodges},\ and\ \citenamefont {Tran}}]{ngo2017}%
  \BibitemOpen
  \bibfield  {author} {\bibinfo {author} {\bibfnamefont {N.}~\bibnamefont {Ngo}}, \bibinfo {author} {\bibfnamefont {H.}~\bibnamefont {Lin}}, \bibinfo {author} {\bibfnamefont {J.}~\bibnamefont {Hodges}}, \ and\ \bibinfo {author} {\bibfnamefont {H.}~\bibnamefont {Tran}},\ }\bibfield  {title} {\enquote {\bibinfo {title} {Spectral shapes of rovibrational lines of co broadened by he, ar, kr and sf$_6$: A test case of the hartmann-tran profile},}\ }\href@noop {} {\bibfield  {journal} {\bibinfo  {journal} {Journal of Quantitative Spectroscopy and Radiative Transfer}\ }\textbf {\bibinfo {volume} {203}},\ \bibinfo {pages} {325--333} (\bibinfo {year} {2017})}\BibitemShut {NoStop}%
\bibitem [{\citenamefont {Tretyakov}\ \emph {et~al.}(2023)\citenamefont {Tretyakov}, \citenamefont {Serov}, \citenamefont {Makarov}, \citenamefont {Vilkov}, \citenamefont {Golubiatnikov}, \citenamefont {Galanina}, \citenamefont {Koshelev}, \citenamefont {Balashov}, \citenamefont {Simonova},\ and\ \citenamefont {Thibault}}]{tretyakov2023}%
  \BibitemOpen
  \bibfield  {author} {\bibinfo {author} {\bibfnamefont {M.~Y.}\ \bibnamefont {Tretyakov}}, \bibinfo {author} {\bibfnamefont {E.}~\bibnamefont {Serov}}, \bibinfo {author} {\bibfnamefont {D.}~\bibnamefont {Makarov}}, \bibinfo {author} {\bibfnamefont {I.}~\bibnamefont {Vilkov}}, \bibinfo {author} {\bibfnamefont {G.~Y.}\ \bibnamefont {Golubiatnikov}}, \bibinfo {author} {\bibfnamefont {T.}~\bibnamefont {Galanina}}, \bibinfo {author} {\bibfnamefont {M.}~\bibnamefont {Koshelev}}, \bibinfo {author} {\bibfnamefont {A.}~\bibnamefont {Balashov}}, \bibinfo {author} {\bibfnamefont {A.}~\bibnamefont {Simonova}}, \ and\ \bibinfo {author} {\bibfnamefont {F.}~\bibnamefont {Thibault}},\ }\bibfield  {title} {\enquote {\bibinfo {title} {Pure rotational r(0) and r(1) lines of co in ar baths: experimental broadening, shifting and mixing parameters in a wide pressure range versus ab initio calculations},}\ }\href@noop {} {\bibfield  {journal} {\bibinfo  {journal} {Physical Chemistry Chemical Physics}\ }\textbf {\bibinfo {volume}
  {25}},\ \bibinfo {pages} {1310--1330} (\bibinfo {year} {2023})}\BibitemShut {NoStop}%
\bibitem [{\citenamefont {Serov}\ \emph {et~al.}(2021)\citenamefont {Serov}, \citenamefont {Stolarczyk}, \citenamefont {Makarov}, \citenamefont {Vilkov}, \citenamefont {Golubiatnikov}, \citenamefont {Balashov}, \citenamefont {Koshelev}, \citenamefont {Wcis{\l}o}, \citenamefont {Thibault},\ and\ \citenamefont {Tretyakov}}]{serov2021}%
  \BibitemOpen
  \bibfield  {author} {\bibinfo {author} {\bibfnamefont {E.}~\bibnamefont {Serov}}, \bibinfo {author} {\bibfnamefont {N.}~\bibnamefont {Stolarczyk}}, \bibinfo {author} {\bibfnamefont {D.}~\bibnamefont {Makarov}}, \bibinfo {author} {\bibfnamefont {I.}~\bibnamefont {Vilkov}}, \bibinfo {author} {\bibfnamefont {G.~Y.}\ \bibnamefont {Golubiatnikov}}, \bibinfo {author} {\bibfnamefont {A.}~\bibnamefont {Balashov}}, \bibinfo {author} {\bibfnamefont {M.}~\bibnamefont {Koshelev}}, \bibinfo {author} {\bibfnamefont {P.}~\bibnamefont {Wcis{\l}o}}, \bibinfo {author} {\bibfnamefont {F.}~\bibnamefont {Thibault}}, \ and\ \bibinfo {author} {\bibfnamefont {M.~Y.}\ \bibnamefont {Tretyakov}},\ }\bibfield  {title} {\enquote {\bibinfo {title} {Co-ar collisions: ab initio model matches experimental spectra at a sub percent level over a wide pressure range},}\ }\href@noop {} {\bibfield  {journal} {\bibinfo  {journal} {Journal of Quantitative Spectroscopy and Radiative Transfer}\ }\textbf {\bibinfo {volume} {272}},\ \bibinfo {pages}
  {107807} (\bibinfo {year} {2021})}\BibitemShut {NoStop}%
\bibitem [{\citenamefont {S{\l}owi{\'n}ski}\ \emph {et~al.}(2022)\citenamefont {S{\l}owi{\'n}ski}, \citenamefont {Makowski}, \citenamefont {So{\l}tys}, \citenamefont {Stankiewicz}, \citenamefont {W{\'o}jtewicz}, \citenamefont {Lisak}, \citenamefont {Piwi{\'n}ski},\ and\ \citenamefont {Wcis{\l}o}}]{slowinski2022}%
  \BibitemOpen
  \bibfield  {author} {\bibinfo {author} {\bibfnamefont {M.}~\bibnamefont {S{\l}owi{\'n}ski}}, \bibinfo {author} {\bibfnamefont {M.}~\bibnamefont {Makowski}}, \bibinfo {author} {\bibfnamefont {K.~L.}\ \bibnamefont {So{\l}tys}}, \bibinfo {author} {\bibfnamefont {K.}~\bibnamefont {Stankiewicz}}, \bibinfo {author} {\bibfnamefont {S.}~\bibnamefont {W{\'o}jtewicz}}, \bibinfo {author} {\bibfnamefont {D.}~\bibnamefont {Lisak}}, \bibinfo {author} {\bibfnamefont {M.}~\bibnamefont {Piwi{\'n}ski}}, \ and\ \bibinfo {author} {\bibfnamefont {P.}~\bibnamefont {Wcis{\l}o}},\ }\bibfield  {title} {\enquote {\bibinfo {title} {Cryogenic mirror position actuator for spectroscopic applications},}\ }\href@noop {} {\bibfield  {journal} {\bibinfo  {journal} {Review of Scientific Instruments}\ }\textbf {\bibinfo {volume} {93}} (\bibinfo {year} {2022})}\BibitemShut {NoStop}%
\bibitem [{\citenamefont {Hartmann}\ \emph {et~al.}(2018)\citenamefont {Hartmann}, \citenamefont {Tran}, \citenamefont {Armante}, \citenamefont {Boulet}, \citenamefont {Campargue}, \citenamefont {Forget}, \citenamefont {Gianfrani}, \citenamefont {Gordon}, \citenamefont {Guerlet}, \citenamefont {Gustafsson}, \citenamefont {Hodges}, \citenamefont {Kassi}, \citenamefont {Lisak}, \citenamefont {Thibault},\ and\ \citenamefont {Toon}}]{hartmann2018}%
  \BibitemOpen
  \bibfield  {author} {\bibinfo {author} {\bibfnamefont {J.-M.}\ \bibnamefont {Hartmann}}, \bibinfo {author} {\bibfnamefont {H.}~\bibnamefont {Tran}}, \bibinfo {author} {\bibfnamefont {R.}~\bibnamefont {Armante}}, \bibinfo {author} {\bibfnamefont {C.}~\bibnamefont {Boulet}}, \bibinfo {author} {\bibfnamefont {A.}~\bibnamefont {Campargue}}, \bibinfo {author} {\bibfnamefont {F.}~\bibnamefont {Forget}}, \bibinfo {author} {\bibfnamefont {L.}~\bibnamefont {Gianfrani}}, \bibinfo {author} {\bibfnamefont {I.}~\bibnamefont {Gordon}}, \bibinfo {author} {\bibfnamefont {S.}~\bibnamefont {Guerlet}}, \bibinfo {author} {\bibfnamefont {M.}~\bibnamefont {Gustafsson}}, \bibinfo {author} {\bibfnamefont {J.~T.}\ \bibnamefont {Hodges}}, \bibinfo {author} {\bibfnamefont {S.}~\bibnamefont {Kassi}}, \bibinfo {author} {\bibfnamefont {D.}~\bibnamefont {Lisak}}, \bibinfo {author} {\bibfnamefont {F.}~\bibnamefont {Thibault}}, \ and\ \bibinfo {author} {\bibfnamefont {G.~C.}\ \bibnamefont {Toon}},\ }\bibfield  {title} {\enquote {\bibinfo
  {title} {Recent advances in collisional effects on spectra of molecular gases and their practical consequences},}\ }\href {\doibase https://doi.org/10.1016/j.jqsrt.2018.03.016} {\bibfield  {journal} {\bibinfo  {journal} {Journal of Quantitative Spectroscopy and Radiative Transfer}\ }\textbf {\bibinfo {volume} {213}},\ \bibinfo {pages} {178--227} (\bibinfo {year} {2018})}\BibitemShut {NoStop}%
\end{thebibliography}%

\end{document}